\documentclass[%
 reprint,
 amsmath,amssymb,
 aps,
]{revtex4-2}

\usepackage{graphicx}% Include figure files
\usepackage{dcolumn}% Align table columns on decimal point
\usepackage{bm}% bold math
\usepackage{braket}
\usepackage{xcolor}

\begin{document}

\preprint{APS/123-QED}

\title{Diffusing wave spectroscopy: a unified treatment on temporal sampling and speckle ensemble methods}% Force line breaks with \\

\author{Jian Xu}
\email{jxxu@caltech.edu.}
\affiliation{Department of Electrical Engineering, California Institute of Technology, Pasadena, California 91125, USA}
\author{Ali K. Jahromi}%
\affiliation{Department of Electrical Engineering, California Institute of Technology, Pasadena, California 91125, USA}
\author{Changhuei Yang}
\affiliation{Department of Electrical Engineering, California Institute of Technology, Pasadena, California 91125, USA}

\date{\today}% It is always \today, today,

\begin{abstract}
Diffusing wave spectroscopy (DWS) is a well-known set of methods to measure the temporal dynamics of dynamic samples. In DWS, dynamic samples scatter the incident coherent light, and the information of the temporal dynamics is encoded in the scattered light. To record and analyze the light signal, there exist two types of methods – temporal sampling methods and speckle ensemble methods. Temporal sampling methods, including diffuse correlation spectroscopy (DCS), use one or multiple large bandwidth detectors to well sample and analyze the temporal light signal to infer the sample temporal dynamics. Speckle ensemble methods, including speckle visibility spectroscopy (SVS), use a high-pixel-count camera sensor to capture a speckle pattern and use the speckle contrast to infer sample temporal dynamics. In this paper, we theoretically and experimentally demonstrate that the decorrelation time ($\tau$) measurement accuracy or SNR of the two types of methods has a unified and similar fundamental expression based on the number of independent observables (NIO) and the photon flux. Given a time measurement duration, NIO in temporal sampling methods is constrained by the measurement duration, while speckle ensemble methods can outperform by using simultaneous sampling channels to scale up NIO significantly. In the case of optical brain monitoring, the interplay of these factors favors speckle ensemble methods. We illustrate that this important engineering consideration is consistent with the previous research on blood pulsatile flow measurements, where a speckle ensemble method operating at 100-fold lower photon flux than a conventional temporal sampling system can achieve a comparable SNR. 
 
% \begin{description}
% Secondary publications and information retrieval purposes.
% You may use the \texttt{description} environment to structure your abstract;
% use the optional argument of the \verb+\item+ command to give the category of each item. 
% \end{description}
\end{abstract}

\maketitle

%\tableofcontents

\section{Introduction}
Diffusing wave spectroscopy (DWS) \cite{Pine1988,Stetefeld2016}  is a well-established approach that is used to measure the temporal dynamical properties of dynamic samples, such as in vivo blood flow monitoring \cite{Durduran2014}, air turbulence quantification \cite{Ancellet1987} and particle diffusion in liquid solution \cite{Boas1995}. A common experimental setting of DWS is to use a coherent laser source to illuminate the dynamic sample and measure the scattered light. The scattered light forms a dynamic speckle pattern in which the information of the sample dynamics is encoded, therefore the sample temporal dynamics can be inferred by analyzing the intensity of scattered light. Recently, DWS has been applied in biomedical and clinical areas, especially in monitoring cerebral blood flow (CBF) \cite{Qureshi2017,Durduran2014,Durduran2010,Yuan2005,Zhao2018,Dunn2001,Boas2010}. In such applications, researchers typically utilize red or near-infrared light to illuminate the brain through skin, probe the dynamic scattering light that interacts with the brain and analyze the recorded light signal to infer the information of CBF. 

\begin{figure}[hbt!]
\centering
\includegraphics[width=0.45\textwidth]{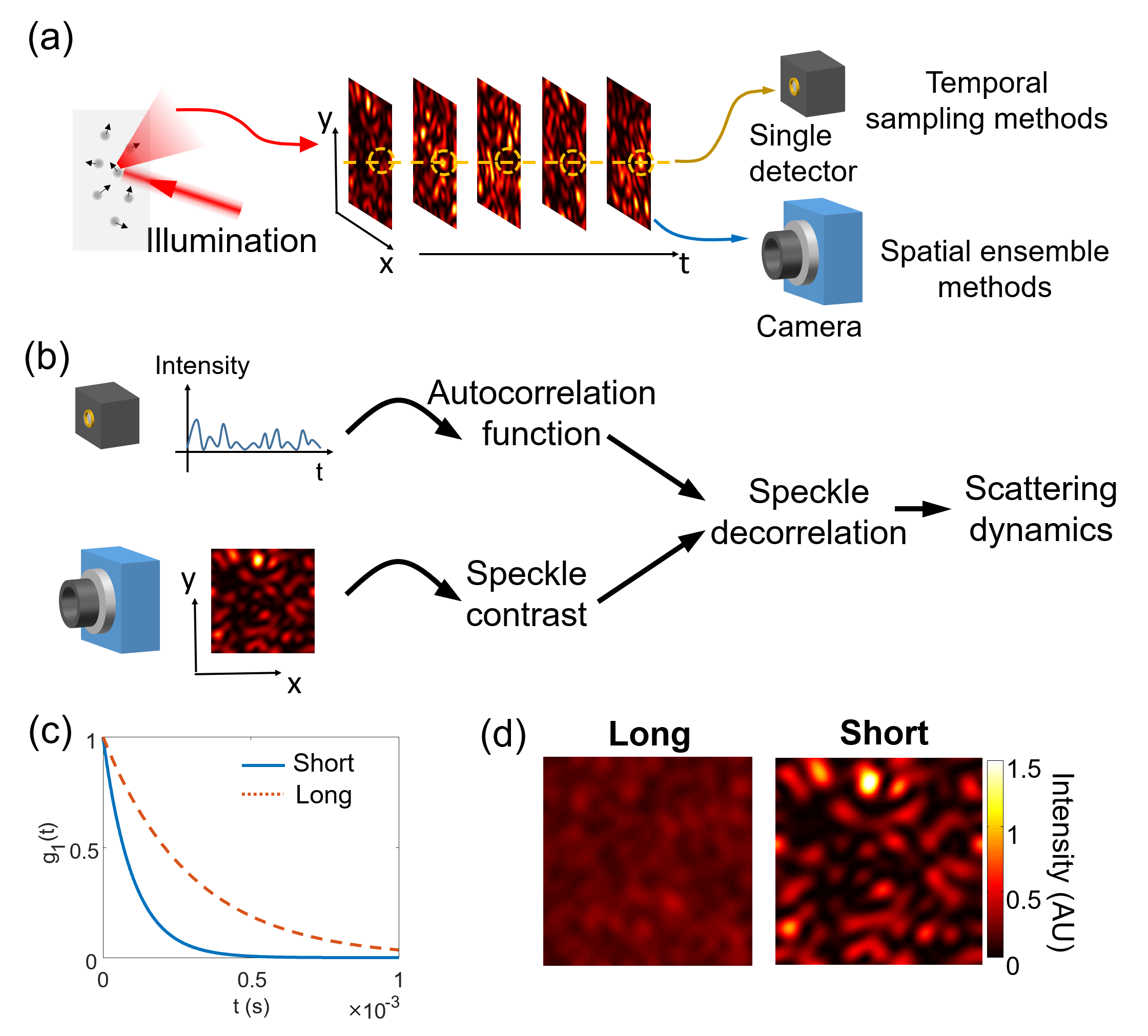}
\caption{\label{DWSfig1} An overview of scattered light dynamics measurement. (a) After the illumination light interacts with the dynamic scatterers, the scattered light forms a set of dynamic speckle patterns. Temporal sampling methods usually use a high speed detector to record the intensity temporal fluctuation, while speckle ensemble methods usually use a camera sensor to record the speckle patterns. (b) Temporal sampling methods calculate the autocorrelation function of the recorded intensity fluctuation to obtain the speckle decorrelation time. Speckle ensemble methods calculate the speckle contrast and use mathematical models to obtain the speckle decorrelation time. In both methods, the calculated speckle decorrelation time is used to infer the scattering dynamics. (c) Examples of field decorrelation functions with a short and a long decorrelation times in temporal sampling methods. (d) Examples of speckle frames with a short and a long decorrelation times in speckle ensemble methods.}

\index{figures}
\end{figure}

Since the dynamic of the light signal is tied to the temporal dynamic of the dynamic sample, there exist two sets of methods to measure the light signal to attain the information of the temporal dynamic – one is to use temporal sampling methods, and the other one is to use speckle ensemble methods. Both methods share similar light illumination systems (Fig. \ref{DWSfig1}a), and the difference is that they collect and analyze the light signal differently.

Temporal sampling methods, including diffuse correlation spectroscopy \cite{Pine1988,Ancellet1987,Boas1995,Durduran2014,Durduran2010}, utilize one or multiple large bandwidth detectors to record the intensity fluctuation of one or a few speckle grains, and analyze the temporal signal to reconstruct the information of the temporal dynamics. The recorded intensity fluctuation trace $I(t)$, where $t$ denotes time, is autocorrelated and normalized to approximate the intensity correlation function $g_2(t)$, i.e., $g_2(t) = \frac{<I(t_1)I(t_1-t)>}{<I(t_1)^2>}$ where $<\cdot>$ denotes the average over time variable $t_1$. According to the Siegert relation \cite{Siegert1943}, the intensity correlation function $g_2(t)$ is $g_2(t) = 1+ |g_1(t)|^2$, where $g_1(t) = \frac{<E(t_1)E(t_1-t)>}{<E(t_1)^2>} $  is the electric field ($E(t)$) correlation function. Speckle decorrelation time is introduced to describe the time scale during which decorrelation happens. Generally, speckle decorrelation time $\tau$ is defined as the time point when the temporal autocorrelation function $g_1(t)$ drops below a certain threshold. A common model is $g_1(t) = \exp(-t/\tau)$ and the time instant that $g_1(t)$ drops to $1/e$ is defined as the decorrelation time. The autocorrelation function of the intensity fluctuation signal can be used to approximate  $g_2(t)$, and it can then be calculated to obtain the speckle decorrelation time and scattering dynamics (Fig. \ref{DWSfig1}b). Figure \ref{DWSfig1}c gives examples of field decorrelation functions with a short decorrelation time and a long decorrelation time.

Typical speckle ensemble methods, including speckle visibility spectroscopy (SVS) \cite{Bandyopadhyay2005,Inoue2012} (a.k.a. speckle contrast spectroscopy \cite{Zhao2018}) and laser speckle contrast imaging (LSCI) \cite{Boas2010,Dunn2011}, use a camera sensor as a detector to record a frame of the speckle pattern. The camera exposure time is longer than the speckle decorrelation time (in experiments, this is generally set at least one order of magnitude longer than the decorrelation time), therefore multiple different speckle patterns sum up within the exposure time, yielding a blurred speckle pattern. The decorrelation time is then calculated from the degree of blurring – more specifically, from the speckle contrast over the speckles in the whole frame.  The speckle contrast $\gamma$ relates with $g_1(t)$ in the form of $\gamma^2 = \int_{0}^{T} 2 (1-\frac{t}{T}) |g_1(t)|^2 dt $  (See Ref. \cite{Bandyopadhyay2005} and Section Appendix \ref{appndix}). From the measured speckle pattern, we can calculate $\gamma$ to obtain $g_1(t)$, consequently obtain information of the sample dynamics (Fig. \ref{DWSfig1}b). Generally, shorter decorrelation time will cause a lower contrast speckle frame. Figure \ref{DWSfig1}d gives examples of speckle frames with a short decorrelation time and a long decorrelation time. 

Since the aforementioned two sets of methods share similar optical illumination but use different detection principles, it is worth exploring the fundamental limitations and jointly analyzing the performance of the two methods. Some previous research have investigated the performance of the two individual methods for several aspects. For instance, ref \cite{Zwanzig1969} discusses the effect of finite sampling time in temporal sampling methods; ref \cite{Koppel1974DWS_noise_original,Zhou2006DWS_noise} build up comprehensive noise models for temporal sampling methods; ref \cite{Inoue2012} discusses the effect of shot noise in speckle ensemble methods. Here, we jointly realize a unified analysis on the performance of the two sets of methods, and show the equivalence of the measurement accuracy of the two methods. Interestingly, we find a unified expression for the two methods with respect to the measurement accuracy. The accuracy of decorrelation time measurements from both sets of methods is determined by the number of independent observables (NIO) and the amount of photon flux. In temporal sampling methods, NIO is the number of decorrelation events recorded by the detector, while in speckle ensemble methods, it is the number of collected speckle grains. The NIO equivalence of the two methods is fundamentally due to the equivalence of spatial speckle ensemble and temporal ensemble.  

Under typical experimental conditions where photon shot noise is the dominant noise source in the measurement, the two sets of methods should provide decorrelation measurements with similar accuracy, given the same NIO and photon flux. In the experiment, we observed that speckle ensemble methods generally have a higher signal-to-noise ratio (SNR) in CBF measurements when the sampling rate is fixed. \textcolor{black}{A speckle ensemble method operating at 100-fold lower photon flux than a conventional temporal sampling method can still achieve a comparable SNR, which is consistent with the results in our previous work \cite{ISVS2020}.} This is because camera sensors used in speckle ensemble methods typically have very large pixel counts, and thereby allow us to achieve a large NIO within the limited measurement time. In contrast, temporal sampling methods, which typically use single-photon-counting-module (SPCM) or other high speed single detectors, tend to lead to a relatively small NIO within the limited measurement time. \textcolor{black}{There have been previous \cite{Dietsche2007AO} and recent \cite{Sie2020FB} efforts in using multiple detectors to boost the effective NIO for temporal sampling methods. However, to date, the number of parallel high-speed detectors deployed in such a fashion is still orders of magnitude lower than the number of pixels available in the standard commercial cameras used in speckle ensemble methods. As we shall explain in the section \ref{Sec-theory}, temporal sampling methods require much more and much faster raw data measurements to generate the same NIO as speckle ensemble methods.}

\section{Theory}\label{Sec-theory}
For the following analysis, we will use optical brain monitoring as the specific reference example. We choose to do this, so that we can map some of the parameters involved into interpretable experimental parameters and promote a better understanding of the factors at play. The analysis itself is general and can be applied to most, if not all, DWS applications. 

\begin{figure}[hbt!]
\centering
\includegraphics[width=0.45\textwidth]{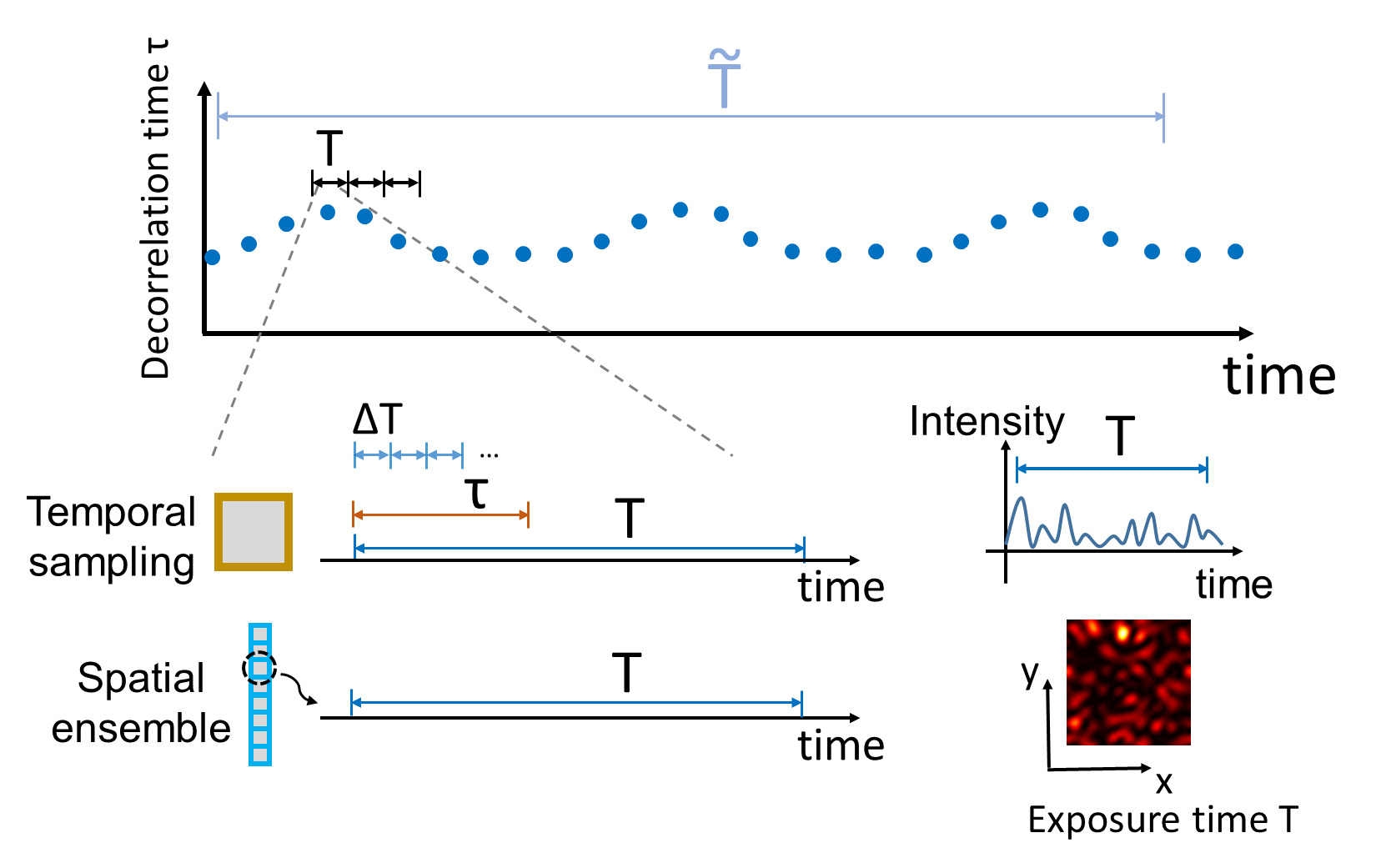}
\caption{\label{DWSfig1.1} An illustration of various time scales defined in the analysis.}

\index{figures}
\end{figure}

To start the quantitative analysis on the two sets of DWS methods, let us first define the various time scales involved in the measurement process (Fig. \ref{DWSfig1.1}). $\Tilde{T}$ denotes the total duration for the measurement process. For a brain monitoring experiment, $\Tilde{T}$ represents the entire duration of the experiment when measurements are made. While it is not used in our subsequent analysis of SNR, we formally define it here so that we are cognizant of this overarching time scale in the entire measurement process. $\tau$ equals the speckle field decorrelation time and it is the quantity that both sets of methods seek to measure. $\tau$ can change over the entire duration of $\Tilde{T}$, and we segment $\Tilde{T}$ into increments of $T$ in order to generate a time trace of $\tau$ measurements (see Fig. \ref{DWSfig1.1} top plot for illustration). $T$ should be chosen so that it is substantially larger than $\tau$ and substantially smaller than the time scale at which $\tau$ is changing. For temporal sampling methods, there is an additional factor $\Delta T$ involved –- $1/\Delta T$ is the rate at which raw intensity measurements are acquired. $\Delta T$ is substantially smaller than $\tau$, as temporal sampling methods require multiple measurements to determine $\tau$. \textcolor{black}{$1/T$ is referred to as the sampling rate, or more specifically, it is the rate at which an estimate of $\tau$ is generated. The terminology can be confusing and $1/T$ should not be confused with $1/\Delta T$, which is the raw data sampling rate in temporal sampling methods. $\Delta T$ is particularly important during system design, as temporal sampling methods require $\Delta T$ to be substantially smaller than $\tau$, so that the intensity fluctuation can be adequately sampled. As $\tau$ in brain monitoring is typically on the order of tens of microseconds, $\Delta T$ needs to be on the order of microseconds or smaller (one order of magnitude smaller than decorrelation time). In turn, this implies that temporal sampling methods require substantially fast detectors.}

Spatially, the two types of DWS methods have similar optimization criterion. In both cases, one should match the detector active pixel area to the typical speckle grain size at the detector plane. In speckle ensemble methods, this may not always be practical. In the event that the speckle grain size is larger than the camera pixel size, we can use the mutual coherence function \cite{Speckle_Goodman} to estimate the speckle grain size. A common parameter of interest for both systems is $N_{\tau}$, a dimensionless number, which is the average number of collected signal photons in one speckle grain per time $\tau$. 

For temporal sampling methods that use a single detector, the SNR of the measured decorrelation time $\tau$, which is defined as the expected decorrelation time $\tau$
divided by $error(\tau)$ (error of  $\tau$ in the measurement), has the form of (Section Appendix \ref{appndix} Eq. \ref{EQDWS_SNR_temporal_2})
\begin{equation}\label{DWSEQt}
    SNR_{temporal} = \frac{\sqrt{2}}{e}\frac{1}{\sqrt{1 + \frac{2}{N_\tau} + \frac{2}{N_\tau^2} \frac{\tau}{\Delta T}}} \sqrt{NIO_{temporal}}.
\end{equation}
Here, $NIO_{temporal}$ is defined as $2T/\tau$, where the constant 2 in $2T/\tau$ is introduced to match the conversion between $g_1(t)$ and $g_2(t)$. Intuitively, $NIO_{temporal}$ is the ratio between the measurement duration $T$ and the decorrelation time $\tau$, which denotes the number of decorrelation events. The detailed derivation is shown in Section Appendix \ref{appndix}.

In speckle ensemble methods, the NIO is equal to the number of independent speckle grains captured by the camera sensor. The SNR of the measured decorrelation time $\tau$ has the form of (Section Appendix \ref{appndix} Eq. \ref{EQDWS_SNR_spatial_2})
\begin{equation}\label{DWSEQs}
\begin{split}
    &SNR_{speckle} = \frac{1}{\sqrt{2}}\frac{1}{1+\frac{1}{N_\tau}} \sqrt{NIO_{speckle}} \\&= \frac{1}{\sqrt{2}} \frac{1}{\sqrt{1+\frac{2}{N_\tau}+\frac{1}{N_\tau^2}}} \sqrt{NIO_{speckle}}.
\end{split}
\end{equation}
$NIO_{speckle}$ is the NIO in speckle ensemble methods. The detailed derivation is shown in Section Appendix \ref{appndix}.

To better interpret this expression, we will briefly describe the measurement system for which this expression would directly apply to. Such a measurement system will have a camera with $NIO_{speckle}$ pixel counts. Each pixel will collect light from a single speckle grain. Each pixel will integrate the collected photons over a time duration of time $T$ and output the result. \textcolor{black}{It is interesting to note that $T$, the camera exposure time in speckle ensemble methods, is not explicitly expressed in Eq. \ref{DWSEQs}. This is because as long as $T$ is substantially longer than $\tau$, a single camera frame capture of the independent speckle grains does not provide any more or less information if $T$ is further lengthened. In brain monitoring experiments, a typical $T$ can be set at $\sim 10\tau$ or longer to ensure 1) the approximation in Eq. \ref{EQDWS_gamma_up_fi} and 2) shot noise dominant detection. Hence, $T$ in speckle ensemble methods is typically more than 100 times larger than $\Delta T$ in temporal sampling methods since typically $\Delta T$ is less than 0.1$\tau$. In turn, this implies that speckle ensemble methods can employ relatively slow commercially-available cameras. }

Equation \ref{DWSEQt} and \ref{DWSEQs} reveals that the two sets of methods have similar dependencies on signal photon counts per speckle per decorrelation time $N_\tau$ and NIO. From a mathematically perspective, we are estimating a statistical parameter (decorrelation time ${\tau}$) from the data, and the accuracy (SNR of ${\tau}$) of the estimation increases with the number of independent sampling points (NIO in this case) according to central limit theorem. In the regime where the photon flux is high enough that $N_{\tau} >> 1$, the center term in both Eq. \ref{DWSEQt} and \ref{DWSEQs} reduces to unity, and both SNR expressions are directly proportional to the square root of NIO. In this regime, the shot noise is negligible compared to the light fluctuations induced by the scatterers’ dynamics. On the other hand, in the regime where $N_{\tau}$ is comparable to or smaller than unity, the center term in both expressions can negatively impact the SNR – the impact of photon shot noise is now more strongly felt. As $\Delta T$ is much smaller than $\tau$, $SNR_{temporal}$ generally is far worse than $SNR_{speckle}$ in this regime. \textcolor{black}{In the grand scheme of things, this factor is relatively minor. Practically, we simply have to make sure the measurements do not operate in this regime.}

Since the SNR “saturates” with respect to $N_{\tau}$ when $N_{\tau} >> 1$, the practical way to perform high accuracy decorrelation time measurements is to increase NIO under the photon sufficient condition ($N_{\tau} >> 1$).

These pair of equations reveal a number of interesting properties for both types of methods. 

In optical brain monitoring, $T$ is constrained as one cannot increase $T$ beyond the time scale of the physiological changes that one is trying to measure. As such, $NIO_{temporal} = 2T/\tau$ has an upper bound. $NIO_{speckle}$ has no such limitation, as NIO is directly dependent on the number of camera pixels that one can use. Ultimately, $NIO_{speckle}$ is constrained by the total area from which we can collect photons, but this limit is seldom reached in optical brain monitoring experiments. For this reason, SNR for speckle ensemble methods can substantially improve over single detector temporal sampling methods by simply increasing camera pixel counts. 

We can also recast the two equations in terms of the amount of measurements made. The total amount of measurements made in $T$ for speckle ensemble methods is $M_{speckle}$ if there are $M_{speckle}$ pixels used to take the speckle frame and each pixel records one speckle grain. From Eq. \ref{DWSEQs}, this leads to an SNR expression:
\begin{equation}
    SNR_{speckle} = \frac{1}{\sqrt{2}} \frac{1}{\sqrt{1+\frac{2}{N_\tau}+\frac{1}{N_\tau^2}}} \sqrt{M_{speckle}}.
\end{equation}
The total amount of measurements made in time $T$ for temporal sampling methods is equal to $M_{temporal} = \frac{2T}{\tau} \times \frac{\tau}{2\Delta T} = NIO_{temporal} \times \frac{\tau}{2\Delta T}$. Substituting it in Eq. \ref{DWSEQt}, this leads to an SNR expression:
\begin{equation}
    \begin{split}
    &SNR_{temporal} \\
    &= \frac{\sqrt{2}}{e}\frac{1}{\sqrt{1 + \frac{2}{N_\tau} + \frac{2}{N_\tau^2} \frac{\tau}{\Delta T}}} \sqrt{M_{temporal}\frac{2\Delta T}{\tau}}.
\end{split}
\end{equation}
We can see that temporal sampling methods require substantially more measurements to achieve the similar SNR as speckle ensemble methods since $\frac{2\Delta T}{\tau}$ is substantially less than unity. 

Yet another way we can interpret the two equations (Eq. \ref{DWSEQs} and \ref{DWSEQt}) is to recast them in terms of the total number of photons collected ($N_{photons}$). For speckle ensemble methods, since $N_{photons} = NIO_{speckle} \times N_{\tau} \times \frac{T}{\tau}$, the SNR expression is given by: 
\begin{equation}
    \begin{split}
    SNR_{speckle} =  \frac{1}{\sqrt{2}} \frac{1}{\sqrt{1+\frac{2}{N_\tau}+\frac{1}{N_\tau^2}}} \sqrt{\frac{N_{photons} }{N_{\tau}} \frac{\tau}{T}}.
\end{split}
\end{equation}

For temporal sampling methods, since $N_{photons} = NIO_{temporal} \times N_{\tau}$, the SNR expression is given by: 
\begin{equation}
    SNR_{temporal} = \frac{\sqrt{2}}{e}\frac{1}{\sqrt{1 + \frac{2}{N_\tau} + \frac{2}{N_\tau^2} \frac{\tau}{\Delta T}}} \sqrt{\frac{N_{photons}}{N_{\tau}}}.
\end{equation}
In the regime where $N_{\tau}$ is $>> 1$, we can see that speckle ensemble methods require substantially more photons to achieve the same SNR as temporal sampling methods. In the context of optical brain monitoring, this situation can occur if 1) the light intensity level is very high so that the condition for $N_{\tau} >> 1$ is met, and 2) both spatial and temporal methods are constrained to only collect the same number of photons. The second condition is highly artificial and can be dismissed, as a well designed speckle ensemble system would try to collect light from as broad an area as possible and, thus, can easily exceed the amount of photons that is collected by a temporal sampling system.  

These two types of equational recasting is helpful because it highlights the impact of the various factors in the SNR expressions for spatial and temporal ensemble methods. 

In Section Appendix \ref{appndix}, we will further examine the SNR expressions for measurement systems that deviate from these designs above. \textcolor{black}{The expression for temporal sampling methods that use multiple parallel detectors is particularly relevant as there have been previous \cite{Dietsche2007AO} and recent \cite{Sie2020FB} efforts focused on such a strategy to improve DWS performance. In brief, such methods can indeed improve SNR. However, they still require raw data measurements at high speed ($\sim $ tens to hundreds of kHz or more). Moreover, such methods still require orders of magnitude more measurements to provide a similar SNR that speckle ensemble methods provide. }

\section{Experiment}
\begin{figure}[hbt!]
\centering
\includegraphics[width=0.45\textwidth]{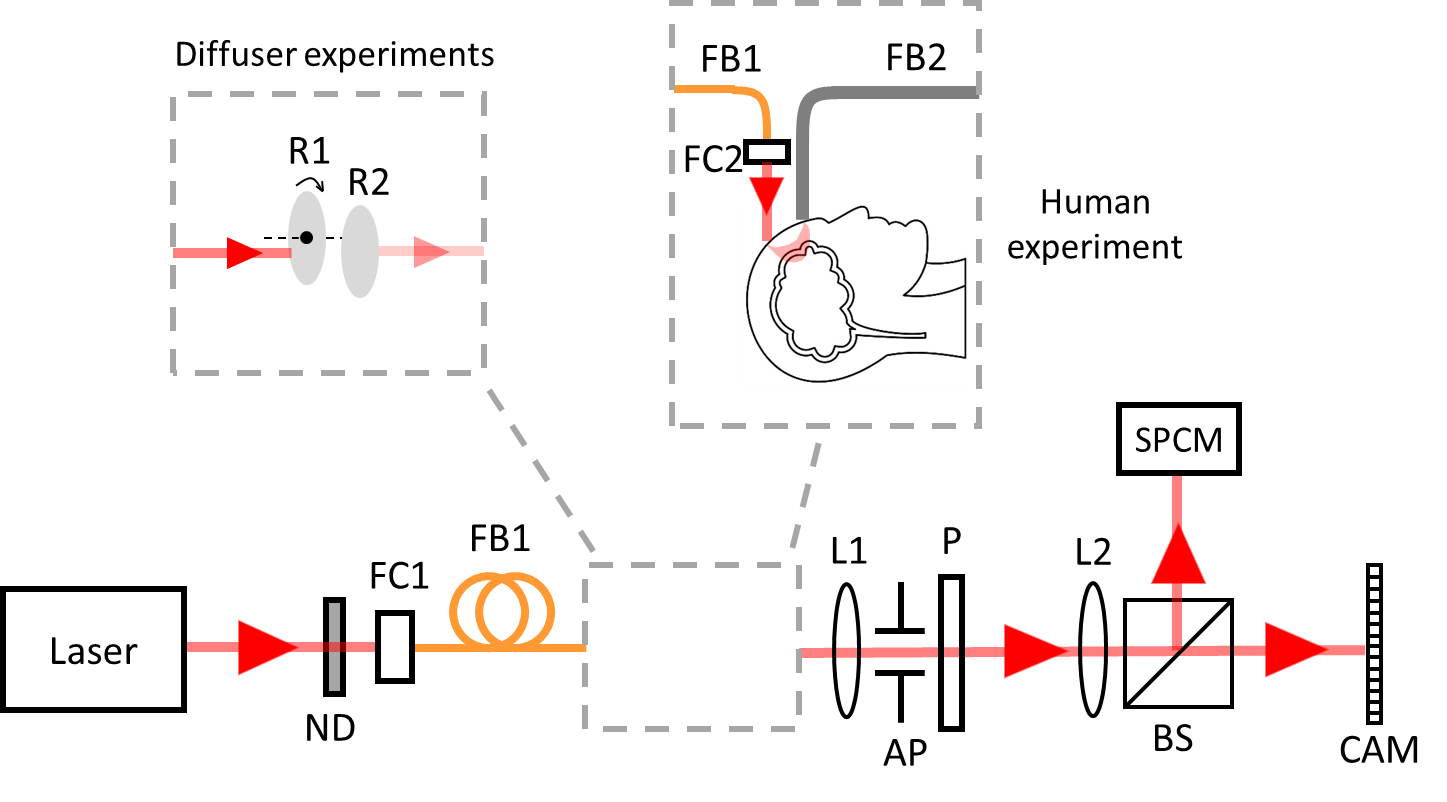}
\caption{\label{DWSfig2} Experimental setup. AP, aperture; BS, beam splitter; CAM, camera; FB, fiber; FC, fiber coupler; L, lens; ND, neutral density filter; P, polarizer; R, rotating diffuser; SPCM, single photon counting module.}
\index{figures}
\end{figure}

We performed experiments to verify the SNR equations (Eq. \ref{DWSEQt} and Eq. \ref{DWSEQs}) of decorrelation time measurements in both temporal sampling and speckle ensemble methods. The experimental setup is shown in Fig. \ref{DWSfig2}. A laser beam (laser model number: CrystaLaser, CL671-150, wavelength 671 nm) is coupled into a multimode fiber FB1, and the output beam from the fiber illuminates the sample (in the gray dashed line box). The scattered light is collected by a 4-f system (L1 and L2), and is split onto a camera and an SPCM (PerkinElmer, SPCM-AQRH-14), respectively. In the diffuser experiment where we verified the models for the two sets of methods, the light passes a rotating diffuser and a static diffuser, and the scattered light is directly collected by the 4-f system. In the human experiment where we demonstrated the NIO advantage of speckle ensemble methods over temporal sampling methods, the light illuminates the skin of the human subject, and diffused light at a source-detector (S-D) separation of 1.5 cm is collected by a large core multimode fiber FB2 (Thorlabs M107L02, core diameter 1.5 mm, containing ~ 6 million modes) and directed to the 4-f system. 

In the human experiment, the 56 mW laser beam with a 6-mm spot size results in a  $< 2 mW/mm^2$ irradiance for skin exposure – within the limit stipulated by American National Standard Institute (ANSI). The output of this fiber was channeled to the camera. A human protocol comprising of all detailed experimental procedures were reviewed and approved by the Caltech Institutional Review Board (IRB) under IRB protocol 19-0941, informed consent was obtained in all cases, and safety precautions were implemented to avoid accidental eye exposure.

\begin{figure}[hbt!]
\centering
\includegraphics[width=0.45\textwidth]{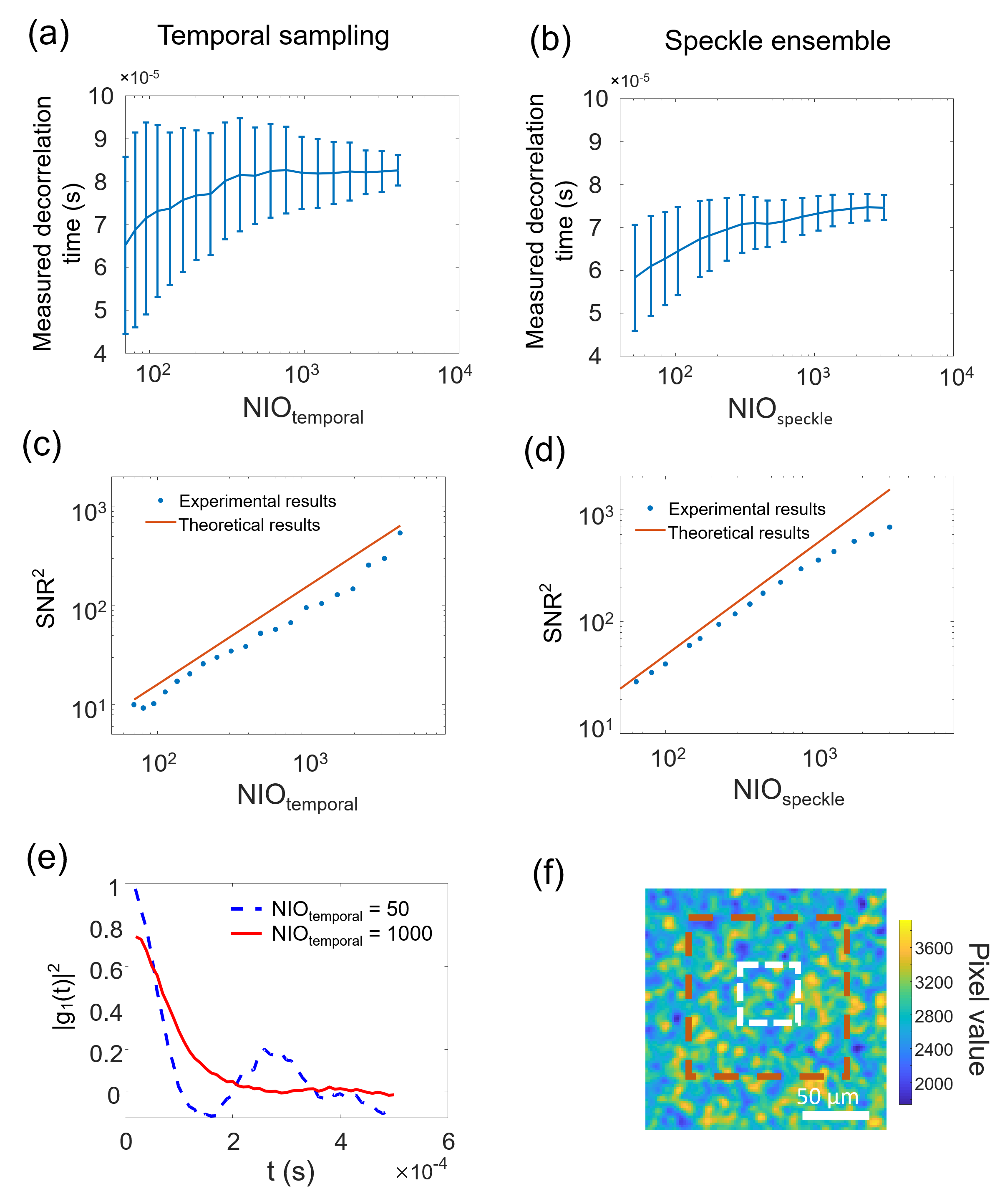}
\caption{\label{DWSfig3} The performance of temporal sampling and speckle ensemble methods with respect to different NIO. (a) Temporal sampling measured speckle decorrelation time with respect to varying NIO. The error bar is calculated from 30 data points. (b) Speckle ensemble measured speckle decorrelation time with respect to varying NIO. The error bar is calculated from 30 data points. (c) The square of SNR with respect to varying NIO in the temporal sampling methods. (d) The square of SNR with respect to varying NIO in the speckle ensemble methods. (e) Examples of the autocorrelation functions from intensity temporal fluctuation traces with different NIO. (f) An examples of a speckle frame used to calculate speckle contrast. The red enclosed box indicates a large $NIO_{speckle}$, and the white enclosed box indicates a small $NIO_{speckle}$.}
\index{figures}
\end{figure}

The experimental results confirm our theoretical analysis. We first verified the relation between SNR and NIO, given a fixed photon flux $N_\tau$. Figure \ref{DWSfig3} shows the relation between the SNR and the NIO under the photon sufficient case, for both temporal sampling and speckle ensemble methods. In the experiment, $N_\tau$ is set to be $\sim10$. In both methods, the square of the SNR scales up linearly with the NIO, as predicted by the theoretical analysis. Due to the approximation in the theoretical analysis (Section Appendix \ref{appndix}) and experimental imperfections such as detector noise and non-perfect control of the diffuser rotating speed, the experimental $SNR^2$ scales up slower compared to the theoretical line. This results in a gap between the experimental dots and the theoretical line in the log-log plot (Fig. \ref{DWSfig3}c,d). The experimentally measured decorrelation times at different NIO are demonstrated in Fig. \ref{DWSfig3}a,b. Both methods can consistently measure the decorrelation time, with less errors as NIO increases. Figure \ref{DWSfig3}e shows examples of the autocorrelation functions from intensity temporal fluctuation traces with different NIO. Under such photon sufficient condition, less NIO will cause the “shape deviation” from the expected autocorrelation function. Intuitively speaking, the number of sampled decorrelation events is not statistically sufficient to be representative for the whole decorrelation process. Figure \ref{DWSfig3}f shows a speckle frame from the speckle ensemble method, with the enclosed red and white boxes containing different number of speckle grains. The speckle contrast calculated from a small enclosed box will give a relatively large error from the expected contrast. Similar as in the temporal sampling method, here in the speckle ensemble method, a small enclosed box does not contain statistically sufficient number of speckle grains to be representative for all the speckle grains in the frame.

\begin{figure}[hbt!]
\centering
\includegraphics[width=0.45\textwidth]{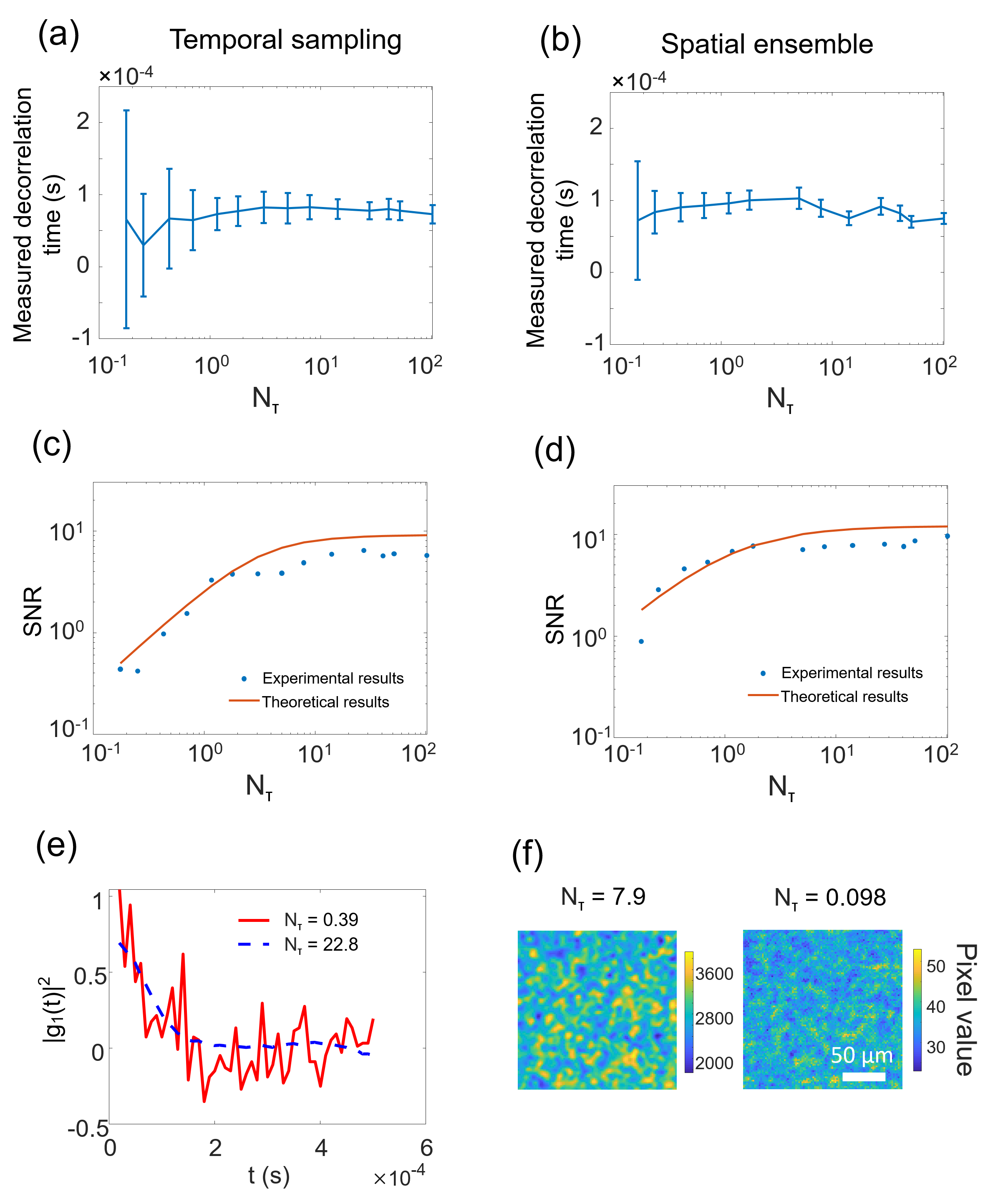}
\caption{\label{DWSfig4} The performance of temporal sampling and speckle ensemble methods with respect to different $N_\tau$. (a) Temporal sampling measured speckle decorrelation time with respect to varying $N_\tau$. The error bar is calculated from 30 data points. (b) Speckle ensemble measured speckle decorrelation time with respect to varying $N_\tau$. (c) SNR with respect to varying $N_\tau$ in the temporal sampling methods. The error bar is calculated from 30 data points. (d) SNR with respect to varying $N_\tau$ in the speckle ensemble methods. (e) Examples of the autocorrelation functions from intensity temporal fluctuation traces with different $N_\tau$. (f) Examples of the speckle frames with different $N_\tau$.}
\index{figures}
\end{figure}

We then verified the relation between SNR and photon flux $N_\tau$, given a fixed NIO. Figure \ref{DWSfig4} shows the relation between the SNR and $N_\tau$ when NIO is set to be 300, for both temporal sampling and speckle ensemble methods. In both methods, the SNR does not change much under the photon sufficient case ($N_\tau>$10), while it starts to decrease when $N_\tau$ is comparable to 1. Figure \ref{DWSfig4}e shows examples of the autocorrelation functions from intensity temporal fluctuation traces with different $N_\tau$. In this case, a small $N_\tau$ will cause more fluctuation in the calculated autocorrelation function. Different from the case of small NIO where the autocorrelation function remains smooth but has a “shape deviation” from the expected autocorrelation function, a small $N_\tau$ here contributes more “noise” on the calculated autocorrelation function. From Eq. \ref{EQDWS_G2tildenoise} and \ref{EQDWS_G2tildevariance} in Section Appendix \ref{appndix}, the fluctuation of the autocorrelation function fundamentally comes from the autocorrelation operation of the noise in the intensity measurement. Figure \ref{DWSfig4}f shows speckle frames from the speckle ensemble method with different $N_\tau$. The low $N_\tau$ speckle frame looks noisier than the high $N_\tau$ frame due to the relatively greater impact of shot noise when $N_\tau$ is low. The shot noise would also contribute to the contrast calculation and subsequently introduces more errors. 

We then implemented both methods to measure human CBF. To well sample the pulsatile effect due to heartbeats, the sampling rate for both methods is set at 18 Hz. The experimental results demonstrate that the speckle ensemble method can reveal the pulsatile effect of the blood flow, while the single channel temporal sampling method cannot. 

\begin{figure}[hbt!]
\centering
\includegraphics[width=0.45\textwidth]{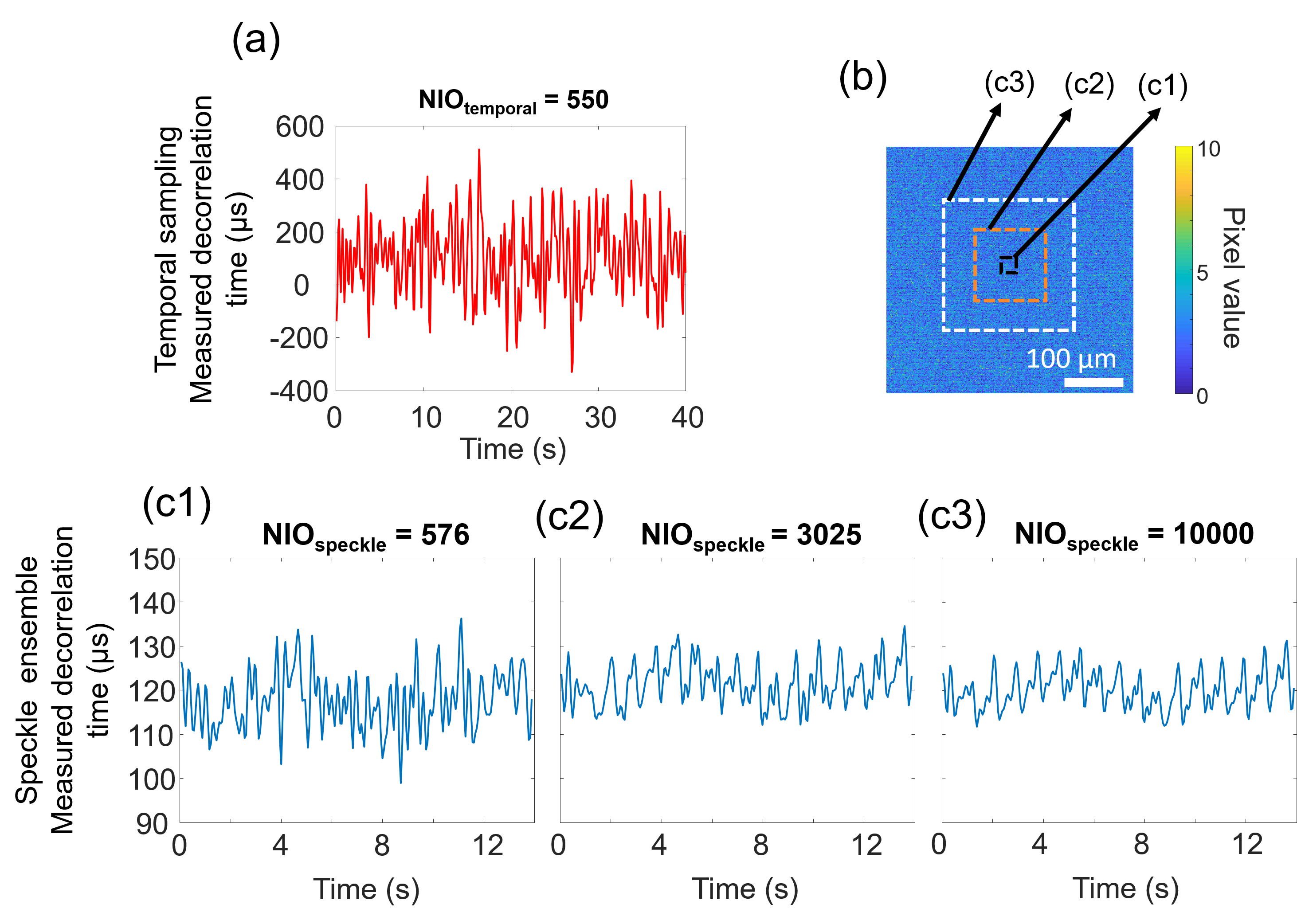}
\caption{\label{DWSfig5} Human CBF induced speckle decorrelation time measurement results from temporal sampling and speckle ensemble methods. (a) Results from the temporal sampling method. (b) A speckle frame from the speckle ensemble method. The white, orange and black boxes enclose 10000, 3025 and 576 speckle grains, respectively. (c) Results from the speckle ensemble method, with different $NIO_{speckle}$.}
\index{figures}
\end{figure}

Under the experimental condition, the photon flux is  $\sim{N}_\tau=0.1$, and the total photon rate is $\sim$1000/(pixel$\cdot$second), which is in the photon starved situation. This photon flux rate is $\sim 100$-fold lower than the operating photon flux in typical DCS experiments. Figure \ref{DWSfig5}a shows the measured decorrelation time of the blood flow by the temporal sampling method (DCS). No obvious pulsatile effect is shown in the plot because of the low measurement SNR. Figure \ref{DWSfig5}c1–\ref{DWSfig5}c3, corresponding to different enclosed boxes in Fig. \ref{DWSfig5}b, show the measured speckle decorrelation time of the blood flow by the speckle ensemble method (SVS) over different number of pixels used in the measurement. In the speckle ensemble method, the measurement SNR increases as the number of pixels used on the camera increases. When the number of pixels is larger than 3025, the pulsatile effect is clearly shown by the speckle ensemble method. 

The reason to the above performance difference between the two sets of methods is tied to the achievable NIO. Under the experimental condition, the photon flux is limited by safety limit. Therefore, a high SNR measurement can only be achieved by a large NIO. In temporal sampling methods, a larger NIO is achieved by measuring more decorrelation events ($2T/\tau$), while in speckle ensemble methods, a larger NIO is achieved by measuring more speckles. Since the sampling rate is fixed to 18 Hz (~56 ms sampling time) and the speckle decorrelation time is mostly determined by CBF (decorrelation time ~0.1 ms), the NIO in temporal sampling is fixed to  ~550. In speckle ensemble methods, increasing the NIO (measuring more speckles) does not affect the sampling rate. In the experiment, the NIO in the speckle ensemble method can achieve more than $10^4$. The difference of achievable NIO between the temporal sampling method and the speckle ensemble method determines the performance difference of the two sets of methods in the speckle decorrelation time measurement. 

In previous temporal sampling methods, to achieve the similar CBF sampling rate with a reasonable measurement SNR, the required photon flux is $\sim$100k/(speckle$\cdot$second) \cite{Wang2016}. In the meantime, costly SPCMs are required to measure the temporal intensity fluctuation. In our demonstrated speckle ensemble method, the photon flux is $\sim$1k/(speckle$\cdot$second), and a common camera sensor is used to measure the diffusing photons. \textcolor{black}{The successful CBF measurement in such a low photon flux condition is also consistent with the results in our previous work \cite{ISVS2020}.} Therefore, the use of a camera sensor relaxes the requirement of the photon budget by two orders of magnitude, and has the potential to allow us to do deeper tissue measurement. 

\section{Discussion}
Our results summarize the performance of the two sets of DWS decorrelation time measurements -- temporal sampling and speckle ensemble methods. We demonstrate that they depend on NIO and photon flux. When $N_\tau >> 1$, i.e. the photon flux is sufficient, the bottle neck of the measurement accuracy is limited by NIO. Since the SNR of the measurement scales up with NIO by a similar constant in the two sets of methods, we can conclude that they have similar “NIO efficiency”, i.e., each independent observable represents “similar amount” of information. However, for one independent observable, temporal sampling methods require ~10 data points or more to construct the decorrelation function, while speckle ensemble methods only require one pixel if we match the speckle size and pixel size. Therefore, speckle ensemble methods can be expected to have higher “data efficiency”.

Based on the current technology, common camera sensors usually support larger data throughputs than common high bandwidth single detectors. Combining with the higher “data efficiency”, speckle ensemble methods should yield more NIO per unit time than temporal sampling methods. Therefore, at the current stage, speckle ensemble methods tend to provide better performance compared to temporal domain methods given the same light illumination and collection architecture. As an example shown in the experimental results, CBF measurement experiment demonstrate the advantage of speckle ensemble methods over temporal sampling methods. Since commercial camera sensors can have millions of pixels, while in Fig. \ref{DWSfig5}c we show that $\sim$3k  pixels are sufficient to monitor the blood flow, there is potential for speckle ensemble methods to do parallel measurements in multiple regions of human brains by using a single camera sensor. 

There is an apparent paradox here in that if the camera exposure time is much longer than the speckle decorrelation time, speckle ensemble methods will have a low contrast, which may be difficult to be measured accurately. However, this paradox in fact does not hold because the SNR expression in Eq. \ref{DWSEQs} is independent on the camera exposure time. In fact, the accuracy of the contrast measurement is mainly determined by the accuracy of the intensity variance measurement (from Section Appendix \ref{appndix} Eq. \ref{EQDWS_gamma_up_hat}), while the mean intensity only scales the intensity variance. Section Appendix \ref{appndix} Eq. \ref{EQDWS_gamma_up_hat} show that the intensity variance measurement only depends on the NIO in the measurement if the camera exposure time is much longer than the speckle decorrelation time. Therefore, the accuracy of the contrast measurement also only depends on the NIO, with no dependency on the contrast value itself. \textcolor{black}{Fundamentally, one can also think that speckle ensemble methods use speckle spatial variance to determine the decorrelation time. Since speckle spatial variance increases as camera exposure time increases, longer exposure time in fact allows camera pixels to better determine the speckle spatial variance. On the other hand, camera exposure time should not exceed the upper limit that causes pixel saturation. }

In general, the analysis of temporal sampling and speckle ensemble methods can be extended to interferometric measurements. In this case, the counterparts of DCS and SVS are IDCS \cite{Zhou2018} and ISVS \cite{ISVS2020}, respectively. We expect that the similar results should also hold in the interferometric schemes, as the mathematical derivations are similar for direct detection discussed in this paper and interferometric detection. 

The drawback of speckle ensemble methods is that they can only provide a measure of the decorrelation time scale, but cannot quantify the exact shape of the decorrelation function. In practice, the combination of the two sets of methods should be able to comprehensively measure the scattering dynamics with high “data efficiency”. Temporal sampling methods can be applied first to quantify the shape of the decorrelation function, while speckle ensemble methods can be applied later to efficiently monitor the relative change of the dynamic scattering. 

\section{Summary}
In conclusion, we performed a systematic analysis on temporal sampling methods and speckle ensemble methods for DWS dynamic scattering measurements. Our theoretical and experimental results demonstrate that the accuracy of two sets of methods is dependent on the number of independent observables and the photon flux. The two sets of methods have similar dependency on the NIO and photon flux. Under the condition where the photon flux is sufficient, the two sets of methods have similar measurement accuracy. We implemented the two sets of methods simultaneously to measure the human CBF, and observed that speckle ensemble methods were able to quantify the CBF with better accuracy than temporal sampling methods, due to higher achievable NIO. We hope our findings can provide the researchers in the field with a guideline of choosing appropriate approaches for dynamic scattering quantification. 

\begin{acknowledgments}
We thank Prof. Yanbei Chen for helpful discussions. This work is supported by the Rosen Bioengineering Center Endowment Fund (9900050).
\end{acknowledgments}

\section{Data Availability}
The data that support the findings of this study are available from the corresponding author upon reasonable request.

\appendix 
\section{Appendix}\label{appndix}
\subsection{SNR of decorrelation time measurements in speckle ensemble methods}
When the light is reflected from the dynamic sample, the light intensity at position $r$ and time $t$, $I_r(t)$, can be decoupled into two parts,
\begin{equation}
    I_r(t) = I_{r,S}(t) + n(t),
\end{equation}
where $I_{r,S}(t)$ is the intensity of one speckle of the signal light that is perturbed by the scattering media, and $n(t)$ is the intensity fluctuation from noise, such as shot noise and detector noise. By the definition of noise, $n(t)$ has zero mean. $I_{r,S}(t)$  follows exponential distribution due to speckle statistics \cite{Speckle_Goodman}. For convenience, we define the AC part of $I_{r,S}(t)$  and $I_{r}(t)$  as $\Tilde{I}_{r,S}(t)$ and $\Tilde{I}_{r}(t)$, respectively, therefore we have
\begin{equation} \label{EQDWS_AC_noise_decom}
    \Tilde{I}_{r}(t) = \Tilde{I}_{r,S}(t) + n(t).
\end{equation}
Here, both $\Tilde{I}_{r,S}(t)$ and $\Tilde{I}_{r}(t)$ are zero mean, and
\begin{equation} \label{EQDWS_speckle_stati}
    \braket{\Tilde{I}_{r,S}(t)} = \sqrt{\braket{\Tilde{I}_{r,S}(t)}} = I_0
\end{equation}
due to speckle statistics \cite{Speckle_Goodman}. Here, $\braket{\cdot}$ denotes the expected value and  $I_0$ is the expected value of $I_{r,S}(t)$. 

Define the signal $S_r$ recorded on the camera pixel located at position $r$ as
\begin{equation}
    S_r = \int_0^T \alpha I_r(t) dt,
\end{equation}
where $\alpha$ is the factor that relates the photon numbers to photon electrons on camera pixels, including detector quantum efficiency, light collection efficiency and other experimental imperfections, $T$ is the camera exposure time.

The speckle contrast $\gamma$ among the whole camera frame is defined as
\begin{equation}
    \gamma = \sqrt{\frac{\braket{(S_r-\braket{S_r})^2}}{\braket{S_r}^2}} \mbox{ or } \gamma^2 = \frac{\braket{(S_r-\braket{S_r})^2}}{\braket{S_r}^2}.
\end{equation}

The numerator of the $\gamma^2$ is
\begin{equation} \label{EQDWS_gammaup}
    \begin{aligned}
    \gamma_{up}^2 &= \braket{(S_r-\braket{S_r})^2} \\
    &= \braket{(\int_0^T \alpha \Tilde{I}_{r}(t) dt)^2} \\
    &= \braket{\int_0^T \int_0^T \alpha^2 \Tilde{I}_{r}(t_1) \Tilde{I}_{r}(t_2) dt_1 dt_2}.
    \end{aligned}
\end{equation}
Substitute equation \ref{EQDWS_AC_noise_decom} to equation \ref{EQDWS_gammaup}, we have 
\begin{widetext}
\begin{equation}
    \begin{aligned}
    \gamma_{up}^2 &= \braket{\int_0^T \int_0^T \alpha^2 \Tilde{I}_{r,S}(t_1) \Tilde{I}_{r,S}(t_2) dt_1 dt_2} + \braket{\int_0^T \int_0^T \alpha^2 n(t_1) n(t_2) dt_1 dt_2} \\
    &= \alpha^2 \braket{\Tilde{I}_r^2} \int_0^T \int_0^T g_S(t_1-t_2) dt_1 dt_2 + \alpha^2 \braket{n^2} \int_0^T \int_0^T g_n(t_1-t_2) dt_1 dt_2 \\
    &= \alpha^2 \braket{\Tilde{I}_r^2} T \int_0^T 2(1-\frac{t}{T}) g_S(t) dt + \alpha^2 \braket{n^2} T \int_0^T 2(1-\frac{t}{T}) g_n(t) dt.
    \end{aligned}
\end{equation}
\end{widetext}
Here, $g_S(t)$ is the correlation function of the mean-removed signal light intensity, and $g_n(t)$ is the correlation function of noise. If we assume $g_S (t) =e^{-2t/\tau}$ and $g_n(t) =e^{-t/\tau_n}$, where $\tau$ is the decorrelation time of the speckle decorrelation time and $\tau_n$ (related to the detector bandwidth BW, $\sim$1/BW) is the noise decorrelation time, in the mean time $T>>\tau$ and $T>>\tau_n$ so that in the integral $1-\frac{t}{T} \approx 1$ before $g_S(t)$ and $g_n(t)$ drop to 0, the above equation can be simplified as
\begin{equation} \label{EQDWS_gamma_up_fi}
    \gamma_{up}^2 \approx \alpha^2 \braket{\Tilde{I}_r^2} T\tau + 2\alpha^2 \braket{n^2} T\tau_n. 
\end{equation}
If the detector is working under the shot noise dominant scheme, where the mean of the number of photon electrons is equal to the standard deviation of the number of photon electrons, we have 
\begin{equation}
    \begin{aligned}
        \alpha I_0 T &= \braket{(\int_0^T \alpha n(t) dt)^2} \\
        &= 2\alpha^2 \braket{n^2} T \tau_n.
    \end{aligned}
\end{equation}
Substitute the above equation and equation \ref{EQDWS_speckle_stati} to equation \ref{EQDWS_gamma_up_fi}, the numerator of the contrast square can be further simplified as
\begin{equation}
    \gamma_{up}^2 \approx \alpha^2 I_0^2 T\tau + \alpha I_0 T.
\end{equation}
The denominator of $\gamma$ is
\begin{equation}
    \gamma_{down} = \braket{S_r} = \alpha I_0 T.
\end{equation}
Hence, the contrast has the expression of
\begin{equation} \label{EQDWS_contrast}
    \begin{aligned}
    \gamma^2 &= \frac{\gamma_{up}^2}{\gamma_{down}^2} \\
    &= \frac{\alpha^2 I_0^2 T\tau + \alpha I_0 T}{(\alpha I_0 T)^2} \\
    &= \frac{\tau}{T} + \frac{1}{\alpha I_0 T} \\
    &= \frac{\tau}{T} + \frac{1}{N_T}
    \end{aligned}
\end{equation}
where $N_T$ is the number of the photon electrons in one speckle within the camera exposure time. Conventional speckle statistics without considering shot noise predicts that the speckle contrast scales with respect to  $1/{\sqrt{N_{pattern}}}$, where $N_{pattern}$ is the number of independent decorrelation patterns recorded by the camera sensor within the exposure time. Intuitively,  $N_{pattern}$  is $\sim T/\tau$  since the ratio provides the number of decorrelation events within the camera exposure time. Here, the extra term $1/\{N_{T}$  in equation \ref{EQDWS_contrast} is due to shot noise, i.e., depending on the photon budget. If the number of photon electrons is sufficient, i.e., $1/\{N_{T} << \tau/T$ , we can discard this term and the expression degenerates to the conventional form. 

In experiment, we can only collect finite number of speckles and use the ensemble average to approximate the contrast. Hence, the contrast square calculated from one camera frame $\hat{\gamma}^2$ is a statistical estimation:
\begin{equation} \label{EQDWS_gamma_finite}
    \hat{\gamma}^2 = \frac{\braket{(S_r-\braket{S_r})^2}_{finite}}{\braket{S_r}_{finite}}.
\end{equation}
Here, $\braket{\cdot}_{finite}$  denotes the ensemble average over the finite speckles in one camera frame. Therefore, both the numerator and denominator of the contrast square $\hat{\gamma}^2$ are estimated from the finite speckles. To evaluate the accuracy of the estimation, we need to estimate the errors of both numerator and denominator in equation \ref{EQDWS_gamma_finite}.

Given a random variable $X$, if we use a sample average $1/N_{independent}\sum_{i=1}^{N_{independent}} X_i$  with $N_{independent}$ independent observables to estimate its expected value  $\braket{X}$, the error between the sample average and the expected value is about $\sqrt{V(X)/N_{independent}}$ , where $V(\cdot)$ denotes the variance of the random variable $X$. In our calculation, $N_{independent}$, the number of independent observables (NIO) in speckle ensemble method, is the number of speckle grains in the camera frame, which is termed $NIO_{speckle}$.

Let us first calculate the variance of the numerator ($\gamma_{up}^2$) of the $\gamma^2$. The variance of $\gamma_{up}^2$ is
\begin{equation} \label{EQDWS_V_gamma}
    \begin{aligned}
    &V(\gamma_{up}^2) = \braket{(S_r-\braket{S_r})^4} - \braket{(S_r-\braket{S_r})^2}^2\\
    &= \alpha^4 \int_0^T \int_0^T \int_0^T \int_0^T \braket{\Tilde{I}_r(t_1)\Tilde{I}_r(t_2)\Tilde{I}_r(t_3)\Tilde{I}_r(t_4)} dt_1 dt_2 dt_3 dt_4 \\
    &\;\;\;\; - \alpha^2 \braket{\int_0^T \int_0^T \Tilde{I}_r(t_1)\Tilde{I}_r(t_2) dt_1 dt_2}.
    \end{aligned}
\end{equation}
The first term in the above equation takes the expected value of four random variables multiplied together. If $\hat{I}_r$ is a Gaussian random variable, the bracket can be expanded as 
\begin{equation}
    \begin{split}
        &\braket{\Tilde{I}_r(t_1)\Tilde{I}_r(t_2)\Tilde{I}_r(t_3)\Tilde{I}_r(t_4)} = \braket{\Tilde{I}_r(t_1)\Tilde{I}_r(t_2)} \braket{\Tilde{I}_r(t_3)\Tilde{I}_r(t_4)} \\
        &+ \braket{\Tilde{I}_r(t_1)\Tilde{I}_r(t_3)} \braket{\Tilde{I}_r(t_2)\Tilde{I}_r(t_4)} + \braket{\Tilde{I}_r(t_1)\Tilde{I}_r(t_4)} \braket{\Tilde{I}_r(t_2)\Tilde{I}_r(t_3)}.
    \end{split}
\end{equation}
Here, even $\Tilde{I}_r$ is not a Gaussian random variable, we still take the formula as an approximation, and this approximation actually holds with tolerable errors based on our experimental results. Equation \ref{EQDWS_V_gamma} then becomes
\begin{widetext}
\begin{equation}
    \begin{split}
    &V(\gamma_{up}^2) \approx \alpha^4 \int_0^T \int_0^T \int_0^T \int_0^T (\braket{\Tilde{I}_r(t_1)\Tilde{I}_r(t_2)} \braket{\Tilde{I}_r(t_3)\Tilde{I}_r(t_4)} \\  
    &+ \braket{\Tilde{I}_r(t_1)\Tilde{I}_r(t_3)} \braket{\Tilde{I}_r(t_2)\Tilde{I}_r(t_4)} + \braket{\Tilde{I}_r(t_1)\Tilde{I}_r(t_4)} \braket{\Tilde{I}_r(t_2)\Tilde{I}_r(t_3)}) dt_1dt_2dt_3dt_4 -\alpha^4 \braket{\int_0^T \int_0^T \Tilde{I}_r(t_1)\Tilde{I}_r(t_2) dt_1 dt_2}^2 \\
    &= 2\alpha^4 \braket{\int_0^T \int_0^T \Tilde{I}_r(t_1)\Tilde{I}_r(t_2) dt_1 dt_2}^2 =2(\gamma_{up}^2)^2.
    \end{split}
\end{equation}
\end{widetext}
Therefore, if there are $NIO_{speckle}$ independent speckles in speckle ensemble methods, the numerator of $\hat{\gamma}_{up}^2$ has a form of 
\begin{equation} \label{EQDWS_gamma_up_hat}
    \begin{split}
        \hat{\gamma}_{up}^2 = {\gamma}_{up}^2 \pm \frac{\sqrt{2}\gamma_{up}^2}{\sqrt{NIO_{speckle}}} = {\gamma}_{up}^2 (1 \pm \sqrt{\frac{2}{NIO_{speckle}}}).
    \end{split}
\end{equation}
Here, the term after the $\pm$ denotes the standard error of the statistical estimation. 

Next, let us calculate the error of the denominator ($\gamma_{down}$) of $\gamma$. It is simply 
\begin{equation}
    \sqrt{\frac{V(S_r)}{NIO_{speckle}}} = \sqrt{\frac{\braket{(S_r-\braket{S_r})^2}}{NIO_{speckle}}} = \sqrt{\frac{\gamma_{up}^2}{NIO_{speckle}}}.
\end{equation}
Therefore, the denominator $\hat{\gamma}_{down}^2$ has a form of 
\begin{equation} \label{EQDWS_gamma_down_hat}
    \begin{split}
        \hat{\gamma}_{down}^2 = (\gamma_{down} \pm \sqrt{\frac{\gamma_{up}^2}{NIO_{speckle}}})^2 \approx \gamma_{down}^2 \pm \frac{2 \gamma_{up} \gamma_{down}}{\sqrt{NIO_{speckle}}}. 
    \end{split}
\end{equation}
Finally, by combining equations \ref{EQDWS_gamma_finite}, \ref{EQDWS_gamma_up_hat} and \ref{EQDWS_gamma_down_hat}, the expression of the estimation of $\hat{\gamma}^2$ is
\begin{equation}
    \begin{split}
        \hat{\gamma}^2 &= \frac{\hat{\gamma}_{up}^2}{\hat{\gamma}_{down}^2} \approx \frac{{\gamma}_{up}^2}{{\gamma}_{down}^2} (1 \pm \sqrt{\frac{1}{NIO_{speckle}}} \sqrt{2+\frac{4\gamma_{up}^2}{\gamma_{down}^2}}) \\
        &=(\frac{\tau}{T}+\frac{1}{N_T}) (1 \pm \sqrt{\frac{1}{NIO_{speckle}}} \sqrt{2+4 (\frac{\tau}{T}+\frac{1}{N_T})}).
    \end{split}
\end{equation}
Hence, the estimation of the contrast is
\begin{equation}
    \begin{split}
        \hat{\gamma} = \sqrt{\frac{\tau}{T}+\frac{1}{N_T}} (1 \pm \frac{1}{2}\sqrt{\frac{1}{NIO_{speckle}}} \sqrt{2+4 (\frac{\tau}{T}+\frac{1}{N_T})}).
    \end{split}
\end{equation}
In SVS, we usually set the camera exposure $T$ much greater than the decorrelation time $\tau$, e.g., $T >> \tau$, and the number of photons collected by one camera pixel $N_T$ is also much greater than 1, e.g., $N_T>>1$. In this case, in the above equation, the second term in the second square root in the error part can be dropped and the estimation of the contrast square $\hat{\gamma}$ can be approximated as
\begin{equation}\label{EQDWS_gamma_hat}
    \begin{split}
        \hat{\gamma} = \sqrt{\frac{\tau}{T}+\frac{1}{N_T}} (1 \pm \sqrt{\frac{1}{2NIO_{speckle}}}).
    \end{split}
\end{equation}
Rewrite the above equation, we have 
\begin{equation}
    \begin{split}
        \tau = T \hat{\gamma}^2 (1 \pm \sqrt{\frac{1}{2NIO_{speckle}}}) -\frac{T}{N_T}
    \end{split}
\end{equation}
The SNR of the decorrelation time in speckle ensemble methods is 
\begin{equation} \label{EQDWS_SNR_spatial1}
    \begin{split}
        SNR_{speckle} &= \frac{\tau}{err(\tau)} = \frac{\tau}{T \hat{\gamma}^2  \sqrt{\frac{1}{2NIO_{speckle}}}} \\
        &= \frac{1}{1+\frac{T}{\tau N_T}} \sqrt{\frac{NIO_{speckle}}{2}}.
    \end{split}
\end{equation}
Here, $err(\tau)$ is the standard error of $\tau$, which is equal to $T \hat{\gamma}^2  \sqrt{\frac{1}{2NIO_{speckle}}}$. Define $N_\tau$ as the number of photon electrons on each camera pixel per time interval $\tau$, we find $N_\tau = \frac{N_T}{T} \tau$. The above equation \ref{EQDWS_SNR_spatial1} can be simplified as 
\begin{equation}\label{EQDWS_SNR_spatial_2}
    \begin{split}
        SNR_{speckle} = \frac{1}{\sqrt{2}}\frac{1}{1+\frac{1}{N_\tau}} \sqrt{NIO_{speckle}}.
    \end{split}
\end{equation}

\subsection{SNR of decorrelation time measurements in temporal sampling methods}

In temporal sampling methods, a fast photodetector with a sufficient bandwidth, such as a single-photon-counting-module (SPCM), is used to well sample the temporal trace $I_r(t)$, and the decorrelation time $\tau$ is computed from the intensity correlation function $G_2(t)$:
\begin{equation}
    G_2(t) = \frac{1}{T}\alpha^2 \int_0^T I_r(t_1) I_r(t_1-t) dt_1.    
\end{equation}
In practice, the correlation is performed between the mean-removed intensity fluctuation:
\begin{equation} \label{EQDWS_G2tilde}
    \Tilde{G}_2(t) = \frac{1}{T}\alpha^2 \int_0^T \Tilde{I}_r(t_1) \Tilde{I}_r(t_1-t) dt_1,   
\end{equation}
where $\Tilde{G}_2(t)$ denotes the intensity correlation function of the two mean-removed intensity traces, $\Tilde{I}_r(t)$ is the AC part of the intensity fluctuation, $t_1$ is the time variable for integral and $t$ is the time offset between the two intensity traces. 

By substituting equation \ref{EQDWS_AC_noise_decom} into equation \ref{EQDWS_G2tilde}, we have
\begin{equation} 
    \Tilde{G}_2(t) = \frac{1}{T}\alpha^2 \int_0^T [\Tilde{I}_{r,S}(t_1)+n(t_1)] [\Tilde{I}_{r,S}(t_1-t)+n(t_1-t)] dt_1.  
\end{equation}
The expected value of $\Tilde{G}_2(t)$ is 
\begin{equation}\label{EQDWS_G2tildenoise}
    \begin{split}
        \braket{\Tilde{G}_2(t)} = \alpha^2 I_0^2 g_S(t) + \alpha^2 \braket{n^2} g_n(t).
    \end{split}
\end{equation}
Same as the definition before, $g_S(t)$ is the correlation function of the mean-removed signal light intensity, and $g_n(t)$ is the correlation function of noise. 

When we use the finite time average to estimate the expected value of $\Tilde{G}_2(t)$, we need to calculate the variance of $\Tilde{G}_2(t)$:
\begin{widetext}
\begin{equation}\label{EQDWS_G2tildevariance}
    \begin{split}
        &V[\Tilde{G}_2(t)] = \frac{1}{T} (\braket{\Tilde{G}_2(t)^2} - \braket{\Tilde{G}_2(t)}^2) \\
        &= \frac{1}{T}\braket{\int_0^T \int_0^T \alpha^4 \Tilde{I}_r(t_1)\Tilde{I}_r(t_1-t) \Tilde{I}_r(t_2)\Tilde{I}_r(t_2-t) dt_1 dt_2} - \frac{1}{T^2} \braket{\Tilde{G}_2(t)^2}^2 \\
        &\approx \frac{2\alpha^4}{T^2} \int_0^T \int_0^T \braket{\Tilde{I}_r(t_1)\Tilde{I}_r(t_2)}^2  dt_1 dt_2 \\
        &= \frac{2\alpha^4}{T^2} \int_0^T \int_0^T [{I_0^2 g_S(t) + \braket{n^2} g_n(t)}]^2  dt_1 dt_2 \\
        & \approx 2(\alpha^4 I_0^4 \frac{\tau}{2T} + \alpha^3 I_0^3 \frac{1}{T} + \alpha^2 I_0^2 \frac{1}{T\tau_n}).
    \end{split}
\end{equation}
\end{widetext}
Hence, if we calculate the correlation function $\Tilde{G}_2(t)$  by using a finite long measurement trace and use it to estimate $\braket{\Tilde{G}_2(t)}$, we have the following estimation form
\begin{widetext}
\begin{equation}
    \begin{split}
        &\Tilde{G}_2(t) = \braket{\Tilde{G}_2(t)} \pm \sqrt{V[\Tilde{G}_2(t)]} \\
        &= [\alpha^2 I_0^2 g_S(t) + \alpha^2\braket{n^2} g_n(t)] \pm \sqrt{2(\alpha^4 I_0^4 \frac{\tau}{2T} + \alpha^3 I_0^3 \frac{1}{T} + \alpha^2 I_0^2 \frac{1}{T\tau_n})}.
    \end{split}
\end{equation}
\end{widetext}
Since $g_n(t)$ usually has much shorter decorrelation time compared to $g_S(t)$, to estimate the speckle decorrelation time $\tau$, we can use the part of the correlation curve where $g_n(t)4$ drops close to 0 while $g_S(t)$ is still close to unity. In this case, the part of the the correlation curve $\hat{G}_2(t)$ is
\begin{equation}
    \begin{split}
        \hat{G}_2(t) &= [\alpha^2 I_0^2 g_S(t)  \pm \sqrt{2(\alpha^4 I_0^4 \frac{\tau}{2T} + \alpha^3 I_0^3 \frac{1}{T} + \alpha^2 I_0^2 \frac{1}{T\tau_n})}\\
        &= \alpha^2 I_0^2 [g_S(t) \pm \sqrt{2(\frac{\tau}{2T} + \frac{1}{\alpha I_0 T} + \frac{1}{\alpha^2 I_0^2T\tau_n})}].
    \end{split}
\end{equation}
In the experiment, $\tau_n$ can be approximated as the inverse of the detector bandwidth, or equivalently the time interval $\Delta T$ between two data points. In the following calculation, we will substitute $\tau_n$ by $\Delta T$.

When we use the decorrelation curve to estimate a parameter associated with the curve, such as decorrelation time, there exist different fitting models to retrieve the parameter. Here, for simplicity, the estimated decorrelation time $\hat{\tau}$ can be chosen by taking the time point where the decorrelation curve drops to $1/e$. In this case, the error of the estimated decorrelation time $err(\tau)$ is 
\begin{equation}
    \begin{split}
        err(\tau) &= \frac{1}{|\frac{dg_S}{dt}|_{g_S(t)=1/e}|}\sqrt{2(\frac{\tau}{2T} + \frac{1}{\alpha I_0 T} + \frac{1}{\alpha^2 I_0^2T\Delta T})}\\
        &=\frac{e}{2}\tau \sqrt{2(\frac{\tau}{2T} + \frac{1}{\alpha I_0 T} + \frac{1}{\alpha^2 I_0^2T\Delta T})}.
    \end{split}
\end{equation}
Hence, the decorrelation time $\tau$ can be estimated from the calculated decorrelation time ${\tau}$ is
\begin{equation}
    \tau = \hat{\tau} (1 \pm \frac{e}{\sqrt{2}}\sqrt{\frac{\tau}{2T} + \frac{1}{\alpha I_0 T} + \frac{1}{\alpha^2 I_0^2T\Delta T}}).
\end{equation}
The SNR of the decorrelation time in temporal sampling methods is 
\begin{equation}\label{EQDWS_SNR_temporal_1}
    SNR_{temporal} = \frac{\tau}{err(\tau)} = \frac{\sqrt{2}}{e}\frac{1}{\sqrt{\frac{\tau}{2T} + \frac{1}{\alpha I_0 T} + \frac{1}{\alpha^2 I_0^2T\Delta T}}}.
\end{equation}
As defined in the main text, the NIO in temporal domain methods $NIO_{temporal} = \frac{2T}{\tau}$, and take the fact that $\alpha I_0 T = \frac{1}{2} NIO_{temporal} N_\tau$, the SNR equation \ref{EQDWS_SNR_temporal_1} can be rewritten as 
\begin{equation}\label{EQDWS_SNR_temporal_2}
    SNR_{temporal} = \frac{\sqrt{2}}{e}\frac{1}{\sqrt{1 + \frac{2}{N_T} + \frac{2}{N_T^2}\frac{\tau}{\Delta T}}}\sqrt{NIO_{temporal}}.
\end{equation}

\subsection{SNR of decorrelation time measurements with other designs}
In this section, we will discuss some experimental designs that deviate from the designs discussed in the main text. For the sake of conciseness, we will give the results with very brief derivation. 

\subsubsection{Temporal sampling: $X$ detectors sampling $X$ independent speckles}
First, let us consider a temporal sampling system where we are able have $X$ separate detectors and are able to measure $X$ independent speckle grains. It is straightforward that the SNR of this system will scale up from Eq. \ref{DWSEQt} by $\sqrt{X}$ times. The SNR of decorrelation time measurements would then become
\begin{equation}\label{DWSappEQt1}
    SNR = \frac{\sqrt{2}}{e}\frac{1}{\sqrt{1 + \frac{2}{N_\tau} + \frac{2}{N_\tau^2} \frac{\tau}{\Delta T}}} \sqrt{NIO_{temporal}} \sqrt{X}.
\end{equation}

\subsubsection{Temporal sampling: one detector sampling $X$ independent speckles}
Second, let us consider a temporal sampling system where we have a single detector but it is made to collect light from $X$ independent speckle grains. In this case, the AC part light intensity $\Tilde{I}_X(t)$ on this detector is 
\begin{equation}\label{EQDWS_one_detector_multi}
    \Tilde{I}_X(t) = \sum_{k=0}^{X} \Tilde{I}_{k}(t) + n_{k}(t),
\end{equation}
where $\Tilde{I}_{k}(t)$ and $n_{k}(t)$ are the $k$-th single speckle intensity and noise, respectively. Following the steps from Eq. \ref{EQDWS_G2tilde} to Eq. \ref{EQDWS_G2tildenoise}, the intensity correlation function $\Tilde{G}_{2,X}(t)$ is 
\begin{equation}
    \Tilde{G}_{2,X}(t) = \frac{1}{T}\alpha^2 \int_0^T \Tilde{I}_X(t_1) \Tilde{I}_X(t_1-t) dt_1,
\end{equation}
and the expected value of $\Tilde{G}_{2,X}(t)$, $\braket{\Tilde{G}_{2,X}(t)}$, is 
\begin{equation}\label{EQDWS_one_detector_multi_EG2}
    \braket{\Tilde{G}_{2,X}(t)} = X \alpha^2 I_0^2 g_S(t) + X \alpha^2 \braket{n^2} g_n(t).
\end{equation}
Following Eq. \ref{EQDWS_G2tildevariance}, the variance of $\Tilde{G}_{2,X}(t)$, $V[\Tilde{G}_{2,X}(t)]$, is 
\begin{equation}\label{EQDWS_one_detector_multi_variance}
\begin{split}
    &V[\Tilde{G}_{2,X}(t)] \approx \frac{2\alpha^4}{T^2} \int_0^T \int_0^T \braket{\Tilde{I}_r(t_1)\Tilde{I}_r(t_2)}^2  dt_1 dt_2 \\
    &= \frac{2\alpha^4}{T^2} \int_0^T \int_0^T [{X I_0^2 g_S(t) + X \braket{n^2} g_n(t)}]^2  dt_1 dt_2 \\
    & \approx 2 X^2 (\alpha^4 I_0^4 \frac{\tau}{2T} + \alpha^3 I_0^3 \frac{1}{T} + \alpha^2 I_0^2 \frac{1}{T\Delta T}).
\end{split}
\end{equation}
The SNR of decorrelation time measurements would then become
\begin{equation}\label{EQDWS_large_detector}
\begin{split}
    &SNR = \frac{\braket{\Tilde{G}_{2,X}(t)}}{\sqrt{V[\Tilde{G}_{2,X}(t)]}}\\
    &= \frac{\sqrt{2}}{e}\frac{1}{\sqrt{1 + \frac{2}{N_\tau} + \frac{2}{N_\tau^2} \frac{\tau}{\Delta T}}} \sqrt{NIO_{temporal}},
\end{split}
\end{equation}
which is the same as Eq. \ref{DWSEQt}.

Paradoxically, a larger detector that collects signal photons from multiple speckles, at first glance, may be expected to yield a decorrelation time measurement with a higher SNR. However,  Eq. \ref{EQDWS_large_detector} implies that a single detector that collects multiple speckles ultimately yields the same SNR as a detector that collects one speckle, if the measurements are shot noise dominant for both cases. 

A mathematically intuitive explanation to this paradox is as followed. When a detector is collecting multiple speckles, the recorded intensity trace is the summation of individual intensity traces of the $X$ collected speckles, as shown in Eq. \ref{EQDWS_one_detector_multi}. Therefore, the expected value of the intensity correlation function scales up with a factor of $X$, as shown in Eq. \ref{EQDWS_one_detector_multi_EG2}. However, during the correlation operation, there are $X^2$ terms, $X$ of which contribute to correlation, while the rest $X(X-1)$ of which contribute to noise. The addition of $X(X-1)$ individual zero-mean random terms scales up the variance term by a factor of ${X(X-1)} \sim X^2$ (shown in Eq. \ref{EQDWS_one_detector_multi_variance}). This intuitive explanation holds when $X$ is large ($X>>1$) and thus ${X(X-1)} \sim X^2$. From the mathematical derivation shown in this subsection, it also holds when $X$ is small. Therefore, the error of the calculated correlation function, which is the square root of the variance, also scales up by a factor of $X$. The simultaneous $X$-fold increase of both the numerator and denominator then cancels each other, and the SNR of decorrelation time measurements does not depend on the number of speckles on the single detector.

Another way to put this is that simply collecting more signal light does not necessarily increase the amount of information or the overall SNR of the system.

\subsubsection{Speckle ensemble: one camera sensor sampling $X$ frames for one decorrelation measurement}
Third, let us consider a speckle ensemble system where instead of putting out a single frame after an exposure time $T$ ($T>>\tau$), it outputs $X$ frames with the same exposure time $T$ for each frame. It is straightforward that the SNR of this system will scale up from Eq. \ref{DWSEQs} by $\sqrt{X}$ times. The SNR equation for this system is given by
\begin{equation}
\begin{split}
    &SNR = \frac{1}{\sqrt{2}} \frac{1}{\sqrt{1+\frac{2}{N_\tau}+\frac{1}{N_\tau^2}}} \sqrt{NIO_{speckle}} \sqrt{X}.
\end{split}
\end{equation}
In fact, as long as the exposure time $T$ is significantly larger than the decorrelation time $\tau$ so that 1) the measurement is shot noise dominant and 2) the approximation in Eq. \ref{EQDWS_gamma_up_fi} holds, increasing $T$ does not improve the decorrelation time measurement accuracy of speckle ensemble methods. Therefore, once a minimal $T$ (empirically 10 times of the decorrelation time $\tau$) satisfies the two conditions, setting the camera exposure time at this $T$ optimizes the overall performance of speckle ensemble methods -- the highest decorrelation time sampling rate with the optimal SNR.

\bibliography{DWS}

\end{document}